
\baselineskip 1cm plus 0pt minus 0pt
\parskip 0pt plus 0pt minus 0pt
\font\vlbf=cmbx10 scaled 1440
\font\lbf=cmbx10 scaled 1200
\hfill\vbox{\halign{&#\hfil\cr &hep-th/9412090\cr &IPM-94-048
                    \cr &IASBS 94-4\cr}}

\hfill{}
\vskip 2cm

\centerline{\vlbf Phase transition in one-dimensional lattice gauge theories}

\vskip 2cm

\centerline{\bf M. Khorrami}

\centerline{\it Institue for Studies in Theoretical Physics and Mathematics}

\vskip -0.5cm\centerline{\it P. O. Box 19395-5746 Tehran, IRAN}
\vskip -0.5cm\centerline{{\it Fax: 98-21-280415
E-mail: mamwad@irearn.bitnet}\footnote*{permanent address}}

\centerline{\it Institue for Advanced Studies in Basic Sciences}

\vskip -0.5cm\centerline{\it P. O. Box 45195-159, Gava Zang, Zanjan, IRAN}

\centerline{\it Department of Physics, Tehran University}

\vskip -0.5cm\centerline{\it North Kargar Ave., Tehran, IRAN}

\vskip 2cm

\centerline {\lbf Abstract }

\noindent Considering one-dimensional nonminimally-coupled lattice gauge
theories, a class of nonlocal one-dimensional systems is presented, which
exhibits a phase transition. It is shown that the transition has a latent
heat, and, therefore, is a first order phase transition.
\vfil\break
\noindent{\lbf O Introduction}

\noindent During the last decades lattice gauge theories have been
extensively studied [1-5]. Lattice theories have no ultraviolet divergences,
they provide a non-perturbative approach to some theories, such as QCD ([1],
for example), and they are theoretically interesting by themselves. They
introduce possibilities, which are absent in the continuum; for example, one
can consider discrete gauge groups as well as continuous ones. So far, the
main interest has been the study of lattice gauge theories (specially
pure-gauge theories) on multidimensional lattices [2,5].

The case of one-dimensional lattices, however, is completely different:
First, one can consider the general form of (minimally-coupled)
gauge-invariant interactions, including matter fields as well as gauge fields
[6]. Second, it is a well-known theorem that one-dimensional systems with
local interactions, do not exhibit phase transition [7]. There are, of
course, examples of nonlocal interactions, which result in phase transition
[7]; one can not, however, deduce them from general principles. Recently,
there has been more interest on one-dimensional systems, with phase
transition [8,9].

Here a class of one-dimensional systems, which is a natural extentions of
minimally-coupled gauge-invariant systems [6], is presented (sections I and
II). Then the analytic behaviour of the free energy of these systems is
considered, and it is shown that, for properly normalized coupling constants,
there is a phase transition (section III); in fact, there is a transition
temperature, above which the pure-gauge interaction of the system
is completely eliminated. Below this temperature, the system goes to a
minimum-energy state of the pure-gauge interaction, and the effects of
nonminimality are lost. Finally, in section IV, the order
of the transition is considered. It is shown that the transition is of first
order, and it has a latent heat. The distinguishing features of this class
of systems is that, first, they provide an example of one-dimensional systems
with phase transition, which arise from a general principle (gauge
invariance) not an artificial modelling, second, the system is easily
solved, and third, the systems belonging to this class possess a certain kind
of universality, that is, the main features of the transition does not depend
on the specific system chosen.

\noindent{\lbf I Nonminimally-coupled lattice gauge theories}

\noindent Consider a lattice consisting of a given set of sites $i$ and
links $<ij>$, two sets $V$ and $\widetilde V$, a function\hskip 3mm
$\widetilde{}\; :\; V\rightarrow\widetilde V$, and a mutiplication from
$\widetilde V\times V$ to $\widetilde V$. The Hamiltonian for a
nearest-neighbour interaction is of the form
$$H_{\circ}:=-\sum_{<ij>}F(\widetilde{S_i}S_j)\eqno ({\rm I}.1)$$
[6,10]; where $F$ is a real-valued function, and $S$ is the matter field (a
$V$-valued function).

Now, suppose that a group $G$ acts on the sets $V$ and $\widetilde V$ through
$$\eqalign{S&\rightarrow \hat g S\cr
\widetilde{(\hat g S)}&=\widetilde S\hat g^{-1},\cr}\eqno({\rm I}.2)$$
where $\hat g$ is a representation of $g$. Introducing a group-element-valued
field defined on links, one reaches a gauge-invariant Hamiltonian
$$H:=H_{\rm m}+H_G,\eqno({\rm I}.3)$$
where
$$H_{\rm m}:=-\sum_{<ij>}F(\widetilde{S_i}\hat U_{<ij>}S_j)
\eqno ({\rm I}.4),$$
and
$$H_G:=-E(W_{l_1},W_{l_2},\dots )\eqno({\rm I}.5).$$
In these definitions, $U$ is a $G$-valued field on links, $E$ is
conjugation-invariant (class) real-valued function of its variables, which
are members of $G$, and the $W_l$'s are Wilson loops of the field U [6].
It is obvious that the Hamiltonian (I.1) is invariant under global gauge
transformation, and (I.3) is invariant under the local gauge tranformation
$$\eqalign{S_i&\rightarrow\hat g_iS_i\cr
 U_{<ij>}&\rightarrow g_iU_{<ij>}g_j^{-1}.\cr}\eqno ({\rm I}.6)$$
This is a minimally-coupled gauge-invariant Hamiltonian [6].

Now, all we need to make $H$ gauge-invariant, is that $U_{<ij>}$ transforms
like (I.6). It need not be a group-valued field. If this field (the gauge
field) is not group-valued, we have a nonminimal coupling.

\noindent{\lbf II One dimensional lattice, and the general form of the
partition function in the thermodynamic limit}

\noindent A one dimensional closed lattice has only one Wilson loop.
So the Hamiltonian (I.3) takes the form
$$H=-\sum_{i=1/2}^{N-1/2} F(\widetilde S_{i-1/2}\hat U_i S_{i+1/2})-E\Big(
\prod\limits_{i=1/2}\limits^{N-1/2} U_i\Big),\eqno ({\rm II}.1)$$
where $N$ is the number of lattice sites,
$$X_{N+k}:=X_k,\eqno({\rm II}.2)$$
and, [6],
$$U_i:=U_{<i-1/2\; i+1/2>}.\eqno({\rm II}.3)$$

Our main goal is to calculate the partition function
$$Z:=\int\Big(\prod\limits_i dS_i\Big)\Big(\prod\limits_j dU_j\Big)\exp\big[
-\beta H({\bf S},{\bf U})\big].\eqno({\rm II}.4)$$
where boldface quantities refer to the set of corresponding quantities on
every site (or link). We also assume that the integration measures are
invariant under the action of group. So, defining a partial
partition function
$$Z_{\rm m}:=\int D{\bf S}\prod\limits_i f(\widetilde S_{i-1/2}\hat U_i
S_{i+1/2}),\eqno({\rm II}.5)$$
where
$$\cases{f:=\exp (\beta F)\cr e:=\exp (\beta E)},\eqno({\rm II}.6)$$
it is easy to see that
$$Z_{\rm m}=\int D{\bf S}\; D{\bf g}\prod\limits_i f(\widetilde S_{i-1/2}
\hat g_{i-1/2}\hat U_i\hat g_{i+1/2}S_{i+1/2}),\eqno({\rm II}.6)$$
where we have normalized the group volume to one.

Now, defining a linear operator $P(U,S,S')$ on the functionals of $G$ through
$$(\psi P)(g):=\int dg'\;\psi (g')f(\widetilde S'\hat g^{\prime -1}\hat U
\hat g S),\eqno({\rm II}.7)$$
one can see that
$$Z_{\rm m}=\int D{\bf S}\;{\rm tr}\Big[\prod\limits_i P(U_i,S_{i-1/2},
S_{i+1/2})\Big].\eqno({\rm II}.8)$$
We want to prove that the eigenvector of $P$, corresponding to its largest
eigenvalue, is independent of its arguments, and that the largest eigenvalue,
itself, depends only on the orbits of the arguments of $P$, provided that
$G$ is compact and the equation
$$Ug=g''U\eqno({\rm II}.9)$$
has always a solution for $g''$.

To establish these properties, we observe that
$$(\psi P)(g)\leq\psi (g_{\rm max})\int dg'\; f(\widetilde S'
\hat g^{\prime -1}\hat U\hat g S),\eqno({\rm II}.10)$$
where $g_{\rm max}$ is the element of group on which $\psi$ attains its
maximum value. This point exists, since the group is compact. Using the
existence of a solution for (II.9), and the invariance of the group measure
under group translations, one can write the above inequality as
$$(\psi P)(g)\leq\psi (g_{\rm max})\int dg'\; f(\widetilde S'
\hat g^{\prime -1}\hat U S).\eqno({\rm II}.11)$$
This inequality also holds when $\psi$ is an eigenvector. So,
$$\lambda\psi (g)\leq\psi (g_{\rm max})\int dg'\; f(\widetilde S'
\hat g^{\prime -1}\hat U S),\eqno({\rm II}.12)$$
where $\lambda$ is the corresponding eigenvalue. In the special case
$g=g_{\rm max}$, one has
$$\lambda\psi (g_{\rm max})\leq\psi (g_{\rm max})\int dg'\;
f(\widetilde S'\hat g^{\prime -1}\hat U S).\eqno({\rm II}.13)$$
One can always make $\psi (g_{\rm max})$ positive. This implies that
$$\lambda\leq\int dg'\; f(\widetilde S'\hat g^{\prime -1}\hat U S).
\eqno({\rm II}.14)$$
It is also seen that the right-hand side of (II.14) is attained for the
constant function. So, the largest eigenvalue of $P$ is
$$\mu (U,S,S'):=\int dg'\; f(\widetilde S'\hat g^{\prime -1}\hat U S).
\eqno({\rm II}.15)$$
Using the existence of a solution for (II.9), and the invariance of the group
measure under group translations, one can also see that
$$\mu (gUg^{\prime -1},\hat g''S,\hat g'''S')=\mu (U,S,S'),
\eqno({\rm II}.16)$$
which is what we wanted to prove.

So, in the thermodynamic limit we have
$$Z=\int D{\bf S}\; D{\bf U}\; e\Big(\prod\limits_i U_i\Big)\Big[
\prod\limits_j\mu\big(\vert U_j\vert ,\vert S_{j+1/2}\vert ,\vert S'_{j-1/2}
\vert\big)\Big],\eqno({\rm II}.17)$$
where {\it absolute value} means the orbit of element under the action of
$G$.

Now, we have
$$\eqalign{\int D{\bf U}\; e\Big(\prod\limits_i U_i\Big)\Big[\prod\limits_j
\mu\big(\vert U_j\vert ,\vert S_{j+1/2}\vert ,\vert S'_{j-1/2}\vert\big)\Big]
&=\int D{\bf U}\; e\Big( g\prod\limits_i U_i\Big)\Big[\prod\limits_j
\mu\big(\vert U_j\vert ,\vert S_{j+1/2}\vert ,\vert S'_{j-1/2}\vert\big)\Big]
\cr &=\int D{\bf U}\; dg\; e\Big( g\prod\limits_i U_i\Big)\Big[\prod\limits_j
\mu\big(\vert U_j\vert ,\vert S_{j+1/2}\vert ,\vert S'_{j-1/2}\vert\big)\Big]
\cr},\eqno({\rm II}.18)$$
Using the existence of a solution in (II.9), it is easy to show that the
integral
$$\nu ({\bf U}):=\int dg\; e\Big( g\prod\limits_i U_i\Big)\eqno({\rm II}.19)
$$
depends only on the orbits of the $U_i$'s; in fact, it depends on the product
of the orbits:
$$\nu ({\bf U})=\nu\Big(\prod\limits_i\vert U_i\vert\Big).\eqno({\rm II}.20)$$
The product on the right-hand side of (II.20) is well-defined, since we have
$$g_1U_1g^{\prime -1}_1g_2U_2g^{\prime -1}_2=g_1g'U_1U_2g^{\prime -1}_2.
\eqno({\rm II}.21)$$
So, one can define the product of two orbits, as the orbit of the product of
two arbitrary elements, one from each orbit.

To conclude, one can write the partition function as
$$Z=\int D{\bf S}\; D{\bf U}\;\nu\Big(\prod\limits_i\vert U_i\vert\Big)
\Big[\prod\limits_j\mu\big(\vert U_j\vert ,\vert S_{j+1/2}\vert ,
\vert S'_{j-1/2}\vert\big)\Big].\eqno({\rm II}.22)$$
This relation holds, provided that the group $G$ is compact, that the
equation (II.9) has always a solution for $g''$, and that the integration
measures are invariant under the action of group.

\noindent{\lbf III A class of one-dimensional sysytems with phase transition}

\noindent Suppose that the matter-field space consists of a single orbit of
the gauge group. It is then easy to show that the partial partition function
$$Z_G:=\int D{\bf U}\;\exp\big[ -\beta H({\bf S},{\bf U})\big]
\eqno({\rm III}.1)$$
does not depend on {\bf S} [6]. This means that one can eliminate the
matter-field from the Hamiltonian, and use a gauge-fixed Hamiltonian
$$H_{\rm gf}:=-\sum_i F^0(U_i)-E\Big(\prod\limits_i U_i\Big),
\eqno({\rm III}.2)$$
where
$$F^0(U):=F(\widetilde S^0\hat US^0),\eqno({\rm III}.3)$$
and $S^0$ is an arbitrary member of the matter-field space. In this case one
has
$$Z=\Big(\int dS\Big)^N\int D{\bf U}\;\exp (-\beta H_{\rm gf}).
\eqno({\rm III}.4)$$

This result holds even for finite lattices. In the thermodynamic limit, using
(II.22), we have
$$Z=\Big(\int dS\Big)^N\int D{\bf U}\;\nu\Big(\prod\limits_i\vert U_i\vert
\Big)\Big[\prod\limits_j\mu\big(\vert U_j\vert\big)\Big].\eqno({\rm III}.5)$$
Now, take a special form for the gauge field: the formal product of a real
number in a set $\{ a_m\}$ and a member of the gauge group:
$$U=vW,\eqno({\rm III}.6)$$
where
$$W\in G,\eqno({\rm III}.7)$$
and
$$v\in\{ a_m\}.\eqno({\rm III}.8)$$
We also assume that the functions $E$ and $F$ are linear with respect to
$v$'s. So we have
$$H_{\rm gf}=-\sum_j v_j F^0(W_j)-\Big(\prod\limits_j v_j\Big)
E\Big(\prod\limits_j W_j\Big).\eqno({\rm III}.9)$$
$E$ is a bounded function, and its bound does not depend on $N$. Therefore
the maximum of $\vert a_m\vert$ should be 1, so that neither $ln Z$ per site
diverges in the thermodynamic limit nor does the pure-gauge part of
interaction disappear in this limit. We now rewrite (III.9) as
$$H_{\rm gf}=-J\sum_j v_j F^0(W_j)-K\Big(\prod\limits_j v_j\Big)
E\Big(\prod\limits_j W_j\Big),\eqno({\rm III}.10)$$
where we have assumed that the maxima of $F^0$ and $E$ in (III.10) is one,
and $\{ a_m\}$ is a subset of $[0,1]$. One can then rewrite (III.5) as
$$Z_{gf}=\sum_{\{ v_j\}}\nu\Big(\prod\limits_i v_i\Big)\Big[\prod\limits_j
\mu (v_j)\Big],\eqno({\rm III}.11)$$
or
$$Z_{gf}=\sum_{\{ v_j\}}I_0^{(E)}\bigg[\beta K\Big(\prod\limits_i v_i\Big)
\bigg]\Big[\prod\limits_j I_0^{(F^0)}(\beta Jv_j)\Big],\eqno({\rm III}.12)$$
where we have defined
$$I_0^{(F)}(x):=\int dg\;\exp\big[ xF(g)\big].\eqno({\rm III}.13)$$
Writing the Taylor series for $I^{(E)}(x)$,
$$I^{(E)}(x)=\sum_{n=0}^\infty\alpha_{0n}^{(E)} x^n,\eqno({\rm III}.14)$$
we will have
$$\eqalign{Z_{\rm gf}&=\sum_{\{ v_j\}}\sum_{n=0}^\infty (\beta K)^n
\alpha_{0n}^{(E)}\prod\limits_j\big[ v_j^n I_0^{(F^0)}(\beta Jv_j)\big]\cr
&=\sum_{n=0}^\infty (\beta K)^n\alpha_{0n}^{(E)}\Big\{\sum_m\big[ a_m^n
I_0^{(F^0)}(\beta Ja_m)\big]\Big\}^N\cr}.\eqno({\rm III}.15)$$
The ratio of different terms of this series varies exponentially with $N$.
So, in the thermodynamic limit only the largest term contributes. We have
then (if $\{ a_m\}\ne\{ 1\}$),
$$\eqalign{Z_{\rm gf}&=\alpha_{00}^{(E)}\Big\{\sum_m\big[ I_0^{(F^0)}
(\beta Ja_m)\big]\Big\}^N\cr&=\alpha_{00}^{(E)}\Big\{\sum_m\big[ I_0^{(F^0)}
(\beta Ja_m)\big]\Big\}^N\cr}.\eqno({\rm III}.16)$$
But in this partition function, there is no trace of the pure-gauge
interaction. One can restore this interaction through renormalizing the
coupling constant:
$$K=:\kappa N^x\eqno({\rm III}.17).$$
We take $\kappa$ to be constant and
$$0<x\leq 1\eqno({\rm III}.18).$$
To make $\ln Z$ per site finite, $x$ should not be greater than 1. One then
has
$$\eqalign{Z_{\rm gf}&=\sum_{n=0}^\infty (\beta\kappa N^x)^n\alpha_{0n}^{(E)}
\Big\{\sum_m\big[ a_m^n I_0^{(F^0)}(\beta Ja_m)\big]\Big\}^N\cr&=:
\sum_n Z_n\cr}.\eqno({\rm III}.19)$$
Now, there is a local maximum for $Z_n$, aside from $Z_0$. Assuming
$n_{\rm max}$ to be large, and using the fact that $I_0^{(E)}(x)$ behaves
like $\exp (x)$ for large $x$, we have
$$\alpha_{0n}^{(E)}\sim{1\over{n!}},\hbox{\hskip 2cm for $n$ large}
\eqno({\rm III}.20)$$
$$\ln Z_n\sim n-\ln n+n\ln (\beta\kappa N^x)+N\ln\big[ I_0^{(F^0)}(\beta J)
\big],\hbox{\hskip 2cm for $n$ large}\eqno({\rm III}.21)$$
which yields
$${{d\ln Z_n}\over{dn}}=\ln{{\beta\kappa N^x}\over n},\eqno({\rm III}.22)$$
or
$$n_{\rm max}=\beta\kappa N^x.\eqno({\rm III}.23)$$
But
$$\ln Z_{n_{\rm max}}=\beta\kappa N^x+N\ln\big[ I_0^{(F^0)}(\beta J)\big]
\eqno({\rm III}.24)$$
is a local maximum. This should be compared with $\ln Z_0$, which is another
local maximum:
$$\ln Z_0=N\ln\Big[\sum_m I_0^{(F^0)}(\beta Ja_m)\Big].\eqno({\rm III}.25)$$
The greater term determines the partition function. But we have
$$\ln{{Z_0}\over{Z_{n_{\rm max}}}}=N\Bigg[\ln{{\sum_mI_0^{(F^0)}(\beta Ja_m)}
\over{I_0^{(F^0)}(\beta J)}}-\beta\kappa N^{x-1}\Bigg].\eqno({\rm III}.26)$$
If $x<1$, this expression is always positive for large $N$; which means that
the pure-gauge interaction is eliminated. However, if $x=1$, there is a
particular value for $\beta$, $\beta_t$, at which this expression changes
sign. So we have
$${1\over N}\ln Z=\cases{\ln\Big[\sum_mI_0^{(F^0)}(\beta Ja_m)\Big],&
$\beta <\beta_t\; (T>T_t)$\cr\noalign{\vskip 1cm}\ln\big[ I_0^{(F^0)}
(\beta J)\big] +\beta\kappa ,&$\beta >\beta_t\; (T<T_t)$.\cr}
\eqno({\rm III}.27)$$
It is seen that above $T_t$, the partition function is independent of
$\kappa$, that is, the system becomes independent of the pure-gauge
interaction. Below $T_t$, the system is independent of $a_m$'s, that is, the
system is freezed in $a_m=1$ and a value for $W_i$'s for which the function
$E\Big(\prod\limits_i W_i\Big)$ is maximum, 1. So, for $T>T_t$, the system
does not see the pure-gauge interaction, whereas for $T<T_t$, the system goes
to the state of minimum energy (of the pure-gauge interaction).

Also note that this renormalization of the coupling constant has a simple
meaning; it means that the pure-gauge interaction introduced in (III.10), is
in fact an interaction density, but a density which is uniform on the
lattice.

\noindent{\lbf IV Order of the transition}

\noindent from (III.27), we have
$${\partial\over{\partial\beta}}\Big({1\over N}\ln Z\Big) =
\cases{J{{\sum_mI_0^{\prime (F^0)}(\beta Ja_m)}\over{
\sum_mI_0^{(F^0)}(\beta Ja_m)}},&
$\beta <\beta_t\; (T>T_t)$\cr\noalign{\vskip 1cm}J{{I_0^{\prime (F^0)}
(\beta J)}\over{I_0^{(F^0)}(\beta J)}}+\kappa ,&$\beta >\beta_t\; (T<T_t)$.
\cr}\eqno({\rm IV}.1)$$
Now,
$$\eqalign{\Delta S&=-\Delta\Big({{\partial A}\over{\partial T}}\Big)\cr
&={\beta\over T}\Delta\Big({{\partial A}\over{\partial\beta}}\Big)\cr
&=-{1\over T}\Delta\Big({{\partial\ln Z}\over{\partial\beta}}\Big)\cr
&={1\over T}\bigg\{ J\Big[{{I_0^{\prime (F^0)}(\beta J)}\over{I_0^{(F^0)}
(\beta J)}}-{{\sum_ma_m I_0^{\prime (F^0)}(\beta Ja_m)}\over
{\sum_m I_0^{\prime (F^0)}(\beta Ja_m)}}\Big] +\kappa\bigg\},\cr}
\eqno({\rm IV}.2)$$
where $S$ and $A$ are the enthropy and the free energy of the system,
respectively. We will see that, a sufficient condition for $\Delta S$ to be
positive is that
$${d\over{dx}}\Big[{{dI_0^{(F^0)}(x)/dx}\over{I_0^{(F^0)}(x)}}\Big] >0.
\eqno({\rm IV}.3)$$
But, for any system, we have
$$C_V={\beta\over T}{{\partial^2 Z}\over{\partial\beta^2}}>0.
\eqno({\rm IV}.4)$$
Using this for a system with $\kappa =0$ and $\{ a_m\} =\{ 1\}$, one can
prove (IV.3). It is now easy to prove that $\Delta S>0$:
$$\Delta S={1\over T}\bigg\{ J{{I_0^{\prime (F^0)}(\beta J)}\over
{I_0^{(F^0)}(\beta J)}}\Big[ 1-{{\sum_mI_0^{\prime (F^0)}(\beta Ja_m)/
I_0^{\prime (F^0)}(\beta J)}\over{\sum_m I_0^{(F^0)}(\beta Ja_m)/
I_0^{(F^0)}(\beta J)}}\Big] +\kappa\bigg\}.\eqno({\rm IV}.5)$$
{}From (IV.3), we have
$${{I_0^{\prime (F^0)}(\beta Ja_m)}\over{I_0^{(F^0)}(\beta Ja_m)}}<
{{I_0^{\prime (F^0)}(\beta J)}\over{I_0^{(F^0)}(\beta J)}},\eqno({\rm IV}.6)$$
or
$${{I_0^{\prime (F^0)}(\beta Ja_m)}\over{I_0^{\prime (F^0)}(\beta J)}}<
{{I_0^{(F^0)}(\beta Ja_m)}\over{I_0^{(F^0)}(\beta J)}}.\eqno({\rm IV}.7)$$
Inserting this inequality in (IV.5), one concludes that
$$\Delta S>0.\eqno({\rm IV}.8)$$
This means that the phase transition has a latent heat. So, it is a
first order phase transition.

\noindent{\lbf Acknowledgement}

\noindent I would like to express my deep gratitude to prof. R. Mansouri for
very useful discussions and encouragement.
\vfil\break
\noindent{\lbf References}

\item{[1]} K. G. Wilson, Phys. Rev. {\bf D10}, No. 8, p 2445, (1974)

\item{[2]} F. J. Wegner, J. Math. Phys. {\bf 12}, No. 10, p 2559, (1971)

\item{[3]} J. B. Kogut, Rev. Mod. Phys. {\bf 51}, No. 4, p 659, (1979)

\item{[4]} R. Balian et. al., Phys. Rev. {\bf D10}, No. 10, p 3376, (1974)

\item{[5]} R. Balian et. al., Phys. Rev. {\bf D11}, No. 8, p 2098, (1975)

\item{[6]} M. Khorrami, ``Exact solution for the most general
minimally-coupled one-dimensional lattice gauge theories'',
Int. J. Theo. Phys., in press

\item{[7]} C. Domb, J. J. Lebowitz, ``Phase transition and critical
phenomena, vol. I'', (1986), Academic press

\item{[8]} M. Mend\`es France, G. Tenenbaum, ``Syst\`emes de points,
diviseurs, et structure fractale'', Bull. Soc. Math. de France, {\bf 121},
(1993)

\item{[9]} M. Mend\`es France, G. Tenenbaum, Commun. Math. Phys. {\bf 154},
p 603, (1993)

\item{[10]} C. Rebbi ed., ``lattice Gauge Theories and Monte Carlo
Simulations'', chapter 2, World Scientific publisher
\end